\documentclass[aps,prd,notitlepage,superscriptaddress,longbibliography,nofootinbib]{revtex4-2}
\usepackage[dvipdfmx]{graphicx}

\usepackage{dcolumn}    
\usepackage{bm}         
\usepackage{hyperref}   
\usepackage{bmpsize}
\usepackage[english]{babel}
\usepackage[skip=1pt, justification=raggedright]{caption}
\usepackage{graphicx}
\vspace{-10pt}
\usepackage{amsmath}
\hypersetup{
    colorlinks=true,
    citecolor=blue,
    linkcolor=blue,
    urlcolor=blue
}

\begin{document}

\title{Science of Cryogenic sub-Hz cROss torsion bar detector with quantum NOn-demolition Speed meter (CHRONOS)
}

\author{Yuki Inoue}
\thanks{Corresponding author: iyuki@ncu.edu.tw}
\affiliation{Department of Physics,National Central University, Taoyuan, Taiwan}
\affiliation{Center for High Energy and High Field (CHiP), National Central University,  Taoyuan, Taiwan}
\affiliation{Institute of Physics, Academia Sinica, Taipei, Taiwan}
\affiliation{Institute of Particle and Nuclear Studies, High Energy Acceleration Research Organization (KEK), Tsukuba, Japan}

\author{Hsiang-Yu Huang}
\affiliation{Department of Physics,National Central University, Taoyuan, Taiwan}
\affiliation{Center for High Energy and High Field (CHiP), National Central University,  Taoyuan, Taiwan}

\author{Vivek Kumar}
\affiliation{Department of Physics,National Central University, Taoyuan, Taiwan}
\affiliation{Center for High Energy and High Field (CHiP), National Central University,  Taoyuan, Taiwan}

\author{Mario Juvenal S. Onglao II}
\affiliation{National Institute of Physics, University of the Philippines - Diliman, Quezon City 1101, Philippines}
\affiliation{Center for High Energy and High Field (CHiP), National Central University,  Taoyuan, Taiwan}

\author{Daiki Tanabe}
\affiliation{Institute of Physics, Academia Sinica, Taipei, Taiwan}
\affiliation{Center for High Energy and High Field (CHiP), National Central University,  Taoyuan, Taiwan}
\affiliation{Institute of Particle and Nuclear Studies, High Energy Acceleration Research Organization (KEK), Tsukuba, Japan}

\author{Ta-Chun Yu}
\affiliation{Department of Physics,National Central University, Taoyuan, Taiwan}
\affiliation{Center for High Energy and High Field (CHiP), National Central University,  Taoyuan, Taiwan}
\date{\today}

\begin{abstract}
The frequency band between $0.1$ and $10\,\mathrm{Hz}$ remains largely unexplored
in gravitational-wave astronomy due to strong seismic,
Newtonian, and suspension thermal noise that limit
ground-based detectors.
The Cryogenic sub-Hz cROss torsion-bar detector with
quantum NOn-demolition Speed meter (CHRONOS)
is a novel detector concept designed to access this
frequency range from the ground.
CHRONOS combines cryogenic torsion-bar test masses
with a triangular Sagnac interferometer implementing
a speed-meter readout, which suppresses
quantum radiation-pressure noise and enables
quantum non-demolition measurements in the sub-Hz regime.
The detector targets a strain sensitivity of
$h \sim 10^{-18}\,\mathrm{Hz^{-1/2}}$
around $2\,\mathrm{Hz}$ and stochastic gravitational wave background of $\Omega_{GW} \sim 2\times 10^{-3}$ at  $2\,\mathrm{Hz}$.
This sensitivity opens a new observational window
between space-based detectors such as LISA
and ground-based interferometers,
enabling observations of intermediate-mass black hole binaries,
searches for stochastic gravitational-wave backgrounds,
and tests of macroscopic quantum measurements.
\end{abstract}

\maketitle
\section{Introduction}
\label{sec:intro}

The direct detection of gravitational waves (GWs) from compact binary mergers
by Advanced LIGO and Advanced Virgo has opened a new observational window
in astronomy and fundamental physics \cite{Abbott2016Discovery}.
KAGRA has further extended the global detector network by introducing
cryogenic interferometry \cite{akutsu2021kagra}.
Current ground-based detectors operate mainly above
$\sim10\,\mathrm{Hz}$,
while the space-based mission LISA will probe the millihertz band
\cite{amaroseoane2017lisa}.

Between these regimes lies the intermediate
$0.1$--$10\,\mathrm{Hz}$ frequency band,
which remains largely unexplored.
Observations in this band from the ground are limited by
seismic noise, Newtonian noise,
and suspension thermal noise.

The Cryogenic sub-Hz cROss torsion-bar detector with quantum NOn-demolition Speed meter (CHRONOS) 
is a novel detector concept designed to explore this frequency range~\cite{inoue2026chronosscienceprogram, inoue2025chronoscryogenicsubhzcross, inoue2026opticaldesignsensitivityoptimization}.
CHRONOS combines cryogenic torsion-bar test masses
with a triangular Sagnac speed-meter interferometer.
The speed-meter configuration suppresses quantum radiation-pressure noise
and enables quantum non-demolition measurements
in the sub-Hz band.

The optical configuration of the detector is illustrated
in Fig.~\ref{fig:optics}.
This configuration allows efficient readout of torsional motion
while maintaining high optical sensitivity.


\section{Detector concept}
\label{sec:concept}

CHRONOS is designed as a ground-based interferometric
gravitational-wave detector optimized for the
sub-Hz frequency band~\cite{inoue2026opticaldesignsensitivityoptimization, inoue2026improvingcalibrationaccuracytorque}.
The detector combines cryogenic torsion-bar test masses
with a triangular Sagnac interferometer implementing
a speed-meter readout.

In conventional gravitational-wave interferometers,
the displacement of the test masses is measured directly.
Such position measurements are fundamentally limited by
quantum radiation-pressure noise at low frequencies,
which leads to the standard quantum limit (SQL)
\cite{Braginsky1992,DanilishinKhalili2012}.
The speed-meter concept provides an alternative measurement
scheme in which the velocity of the test mass is measured
instead of its position.
Because the velocity corresponds to the time derivative
of displacement, quantum back-action noise is strongly
suppressed at low frequencies \cite{Chen2003}.

The speed-meter readout can be realized using a
Sagnac interferometer configuration.
In this topology, the light sequentially probes
the test masses along two counter-propagating paths.
The phase difference between the two beams
is proportional to the velocity of the test mass,
effectively implementing a quantum non-demolition
measurement of the mechanical motion.

The CHRONOS interferometer adopts a triangular Sagnac
configuration with power recycling and signal recycling
cavities, as illustrated in Fig.~\ref{fig:optics}.
The torsion-bar test masses provide extremely low
mechanical resonance frequencies,
which allow the detector to respond efficiently
to gravitational waves in the sub-Hz band.
The angular displacement of the torsion bars
is read out interferometrically,
providing high sensitivity to gravitational-wave signals.

Cryogenic operation plays a crucial role
in suppressing thermal noise.
By cooling the suspension and test masses,
thermal fluctuations in the torsional degree of freedom
are significantly reduced.
This combination of cryogenic torsion-bar mechanics
and speed-meter interferometry enables
a new experimental regime of precision measurement
in the sub-Hz frequency band.

\section{Sensitivity}

\begin{figure}[t]
\centering

\begin{minipage}{0.48\linewidth}
\centering
\includegraphics[width=\linewidth]{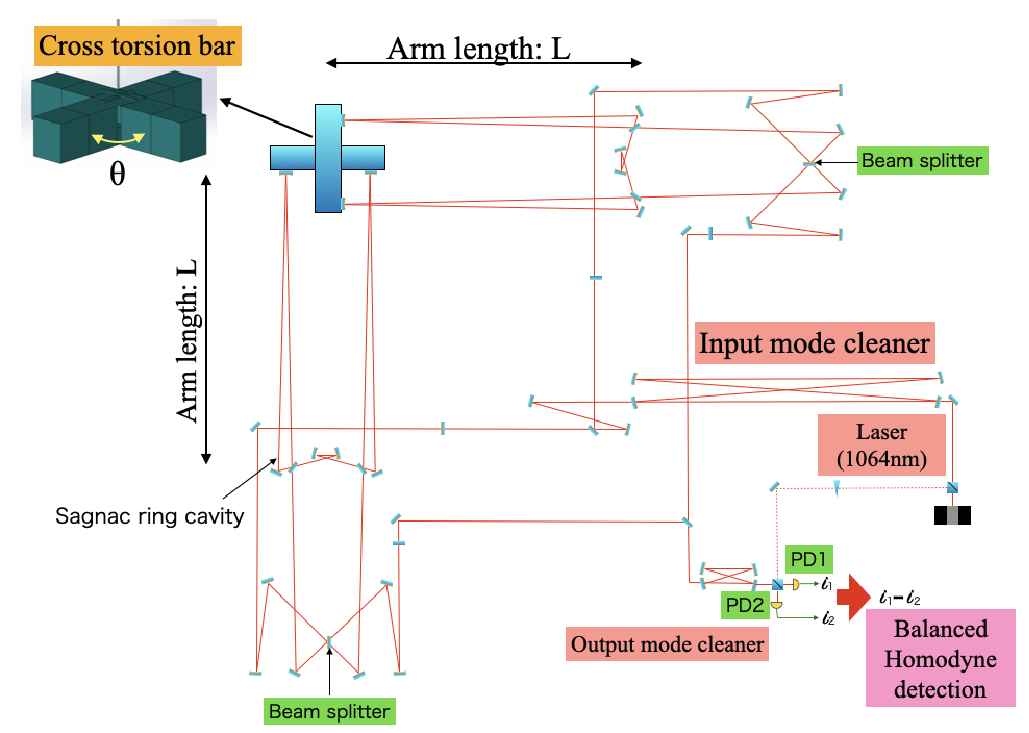}
\caption{Optical configuration of CHRONOS. The detector adopts a triangular Sagnac interferometer with torsion-bar test masses implementing a speed-meter readout.}
\label{fig:optics}
\end{minipage}
\hfill
\begin{minipage}{0.48\linewidth}
\centering
\includegraphics[width=\linewidth]{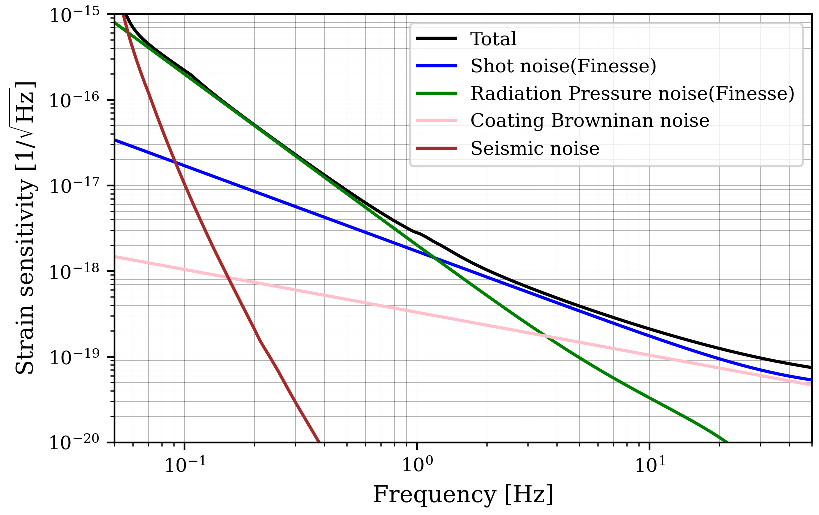}
\caption{Projected strain sensitivity of CHRONOS. The detector is optimized for the sub-Hz frequency band between $0.1$ and $10\,\mathrm{Hz}$.}
\label{fig:noise}
\end{minipage}

\end{figure}

The projected strain sensitivity of CHRONOS
is shown in Fig.~\ref{fig:noise}.
The detector is optimized for the
$0.1$--$10\,\mathrm{Hz}$ frequency band,
which lies between the sensitivity ranges
of space-based detectors such as LISA
and current ground-based interferometers.
Accessing this frequency band from the ground
requires careful mitigation of several
fundamental noise sources that dominate
at low frequencies ~\cite{inoue2026opticaldesignsensitivityoptimization, tanabe2025torquecancellationeffectintensity}.

At frequencies below a few hertz,
seismic motion of the ground represents
one of the primary limitations.
Even with advanced seismic isolation systems,
residual ground motion can couple into the
test-mass motion and degrade the detector sensitivity.
Closely related to seismic motion is
Newtonian noise, also referred to as
gravity-gradient noise,
which arises from time-dependent fluctuations
in the local gravitational field produced by
moving masses in the surrounding environment,
such as seismic waves and atmospheric density variations.
Because Newtonian noise directly modulates
the gravitational field acting on the test masses,
it cannot be shielded and must instead be mitigated
through site selection, environmental monitoring,
and subtraction techniques.

Quantum noise constitutes the fundamental limit
for interferometric gravitational-wave detectors.
In conventional position-meter interferometers,
quantum radiation-pressure noise becomes dominant
at low frequencies due to back-action from
photon momentum fluctuations.
CHRONOS mitigates this limitation by adopting
a speed-meter interferometer configuration,
in which the velocity of the test mass is measured
instead of its displacement.
This measurement scheme suppresses quantum back-action
and allows the detector to approach or surpass
the standard quantum limit in the low-frequency regime.

Combining these design elements,
the expected strain sensitivity of CHRONOS
reaches approximately

\[
h \sim 10^{-18}\,\mathrm{Hz^{-1/2}}
\]

around $f \sim 2\,\mathrm{Hz}$.
This sensitivity represents a significant improvement
in the sub-Hz band and opens a new observational window
for ground-based gravitational-wave detectors.
Such performance enables astrophysical observations
that are inaccessible to both existing ground-based
interferometers and space-based detectors,
thereby bridging the gap between the two regimes.

\begin{figure}[t]
\centering

\begin{minipage}{0.48\linewidth}
\centering
\includegraphics[width=\linewidth]{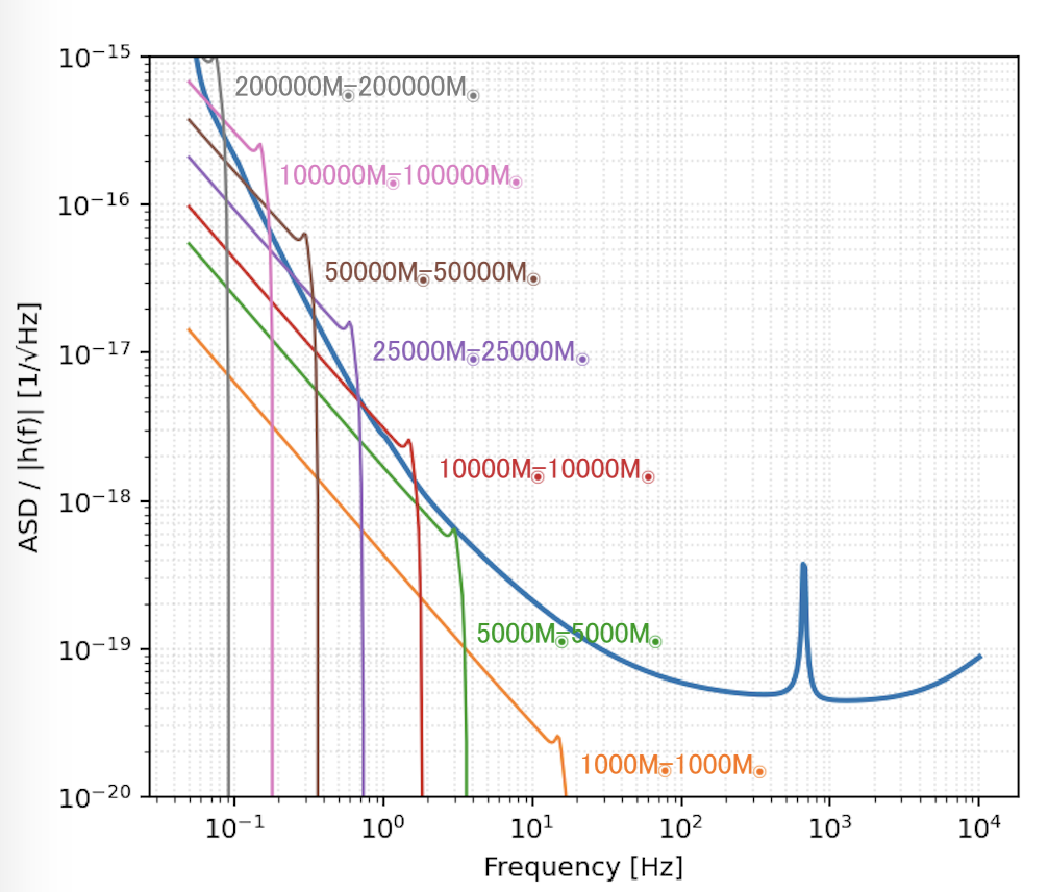}
\caption{Expected sensitivity of CHRONOS to the stochastic gravitational-wave background expressed in terms of the energy density spectrum $\Omega_{\mathrm{GW}}(f)$.}
\label{fig:sgwb}
\end{minipage}
\hfill
\begin{minipage}{0.48\linewidth}
\centering
\includegraphics[width=\linewidth]{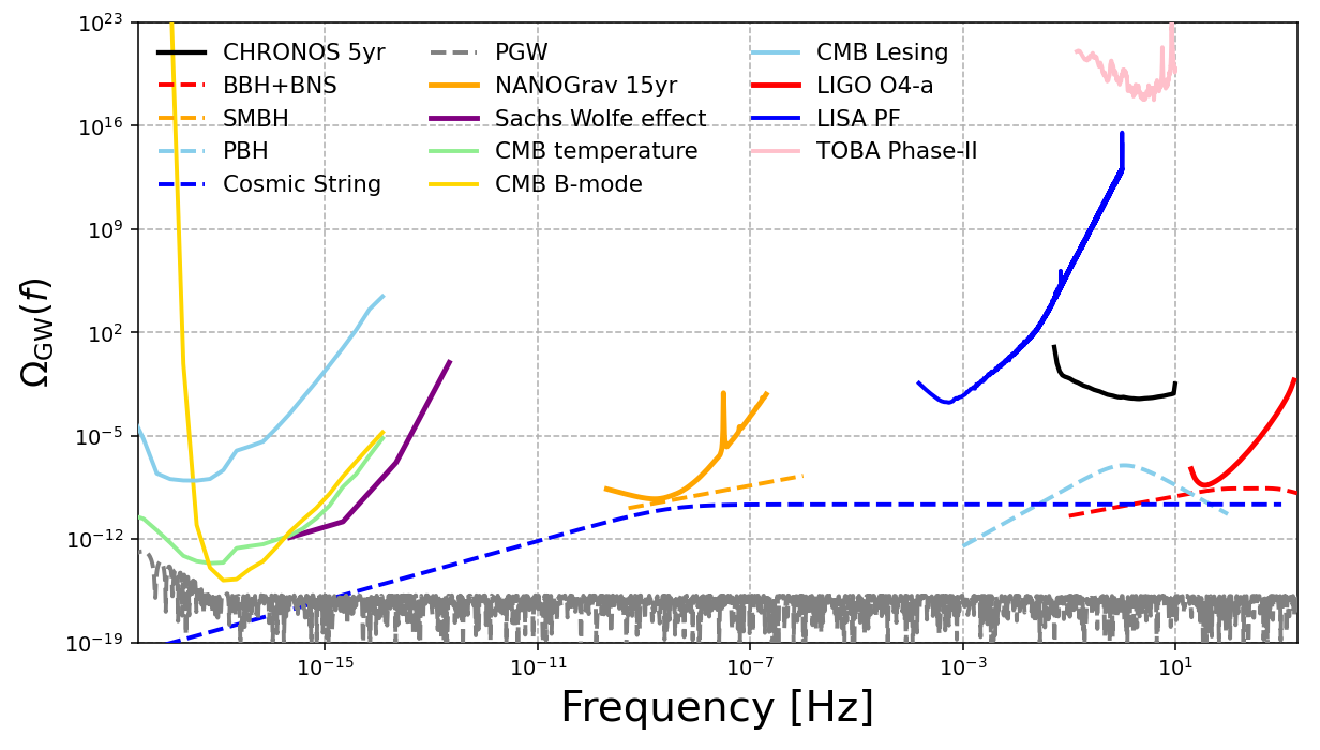}
\caption{Combined constraints on $\Omega_{\mathrm{GW}}(f)$ including CHRONOS, CMB observations, and ground-based interferometers.}
\label{fig:sgwb_combined}
\end{minipage}

\end{figure}
\section{Science}

Observations in the sub-Hz frequency band open a new
scientific domain in gravitational-wave astronomy.
This frequency range lies between the millihertz band
targeted by space-based detectors and the
tens-of-hertz band explored by current ground-based
interferometers.
Gravitational-wave signals in the $0.1$--$10\,\mathrm{Hz}$
band contain crucial information about the early inspiral
phase of compact binaries, the stochastic gravitational-wave
background, and low-frequency gravitational phenomena that
cannot be observed in other frequency ranges.
By accessing this band from the ground,
CHRONOS enables a broad range of astrophysical,
cosmological, and fundamental-physics studies.

\textbf{Intermediate-mass black holes.}
One of the primary science targets of CHRONOS
is the observation of intermediate-mass black hole (IMBH)
binaries with total masses of
$10^2$--$10^4\,M_\odot$ \cite{miller2004imbh}.
While stellar-mass and supermassive black holes
are well established observationally,
the existence and formation pathways of IMBHs
remain uncertain.
The sub-Hz band corresponds to the late inspiral
phase of such systems,
where the gravitational-wave signal evolves slowly
and remains in the detector band for extended durations.
Detecting these signals would provide important insight
into the formation and growth of black holes,
as well as the dynamical processes occurring
in dense stellar environments and galactic nuclei.

\textbf{Stochastic gravitational-wave background.}
CHRONOS also has the potential to probe
the stochastic gravitational-wave background (SGWB)
in the sub-Hz regime.
The SGWB arises from the superposition of many
unresolved astrophysical sources
and may also contain contributions from
processes in the early Universe.
Figure~\ref{fig:sgwb} shows the projected sensitivity
to the gravitational-wave energy density spectrum
$\Omega_{\mathrm{GW}}(f)$.
CHRONOS detector can reach $\Omega_{GW} \sim 2\times 10^{-3}$ at  $2\,\mathrm{Hz}$ with 5 year accumulation time.
Cosmological sources such as
first-order phase transitions
and cosmic strings are predicted to generate
gravitational-wave spectra that may peak
in the sub-Hz band
\cite{caprini2018cosmological}.
By extending observations into this frequency range,
CHRONOS complements both cosmic microwave background
constraints at extremely low frequencies
and ground-based interferometer limits
at higher frequencies.

\textbf{Quantum measurement.}
Beyond astrophysical observations,
CHRONOS provides a unique opportunity to explore
fundamental aspects of quantum measurement.
The sensitivity of interferometric detectors
is fundamentally limited by quantum noise.
The speed-meter configuration adopted by CHRONOS
suppresses quantum radiation-pressure noise
by measuring the velocity of the test masses
rather than their displacement
\cite{Chen2003}.
This approach enables measurements that
can surpass the standard quantum limit
\cite{Braginsky1992}.
Demonstrating quantum non-demolition measurements
in the sub-Hz band would represent
a major milestone in macroscopic quantum physics
and precision interferometry.

\textbf{Prompt gravity signals from earthquakes.}
Finally, CHRONOS may also detect prompt gravity signals
associated with large earthquakes.
Rapid redistribution of mass during seismic events
produces changes in the Earth's gravitational field,
which propagate at the speed of light.
These gravity perturbations arrive at distant locations
earlier than seismic waves and therefore provide
a potential method for early earthquake detection
\cite{Montagner2016}.
Observations of such signals would also improve
our understanding of gravity-gradient noise,
which is a critical limitation for low-frequency
gravitational-wave detectors.

Taken together, these scientific objectives illustrate
the broad impact of CHRONOS across multiple disciplines,
including gravitational-wave astronomy,
cosmology, quantum measurement,
and geophysics.


\section{Conclusions}

CHRONOS is a next-generation
ground-based gravitational-wave detector
designed to explore the
$0.1$--$10\,\mathrm{Hz}$ band.

By combining cryogenic torsion-bar test masses
with a triangular Sagnac speed-meter interferometer,
CHRONOS suppresses quantum radiation-pressure noise
and enables quantum non-demolition measurements
in the sub-Hz regime.

With a projected strain sensitivity of
$h \sim 10^{-18}\,\mathrm{Hz^{-1/2}}$
at $2\,\mathrm{Hz}$,
CHRONOS opens a new observational window
between LISA and terrestrial interferometers.
The detector enables observations of
intermediate-mass black holes,
stochastic gravitational-wave backgrounds,
and tests of macroscopic quantum mechanics,
while also providing potential applications
in geophysics.

CHRONOS therefore represents a key step toward
future multi-band gravitational-wave astronomy.

\begin{acknowledgments}
We would like to express our sincere gratitude to Y-C.Lin, A.Markoshan, M.Hasegawa, T.Kanayama and M.Hazumi for their valuable discussions and continuous support throughout this work.
We also acknowledge the support and collaborative environment provided by the Department of Physics and the Center for High Energy and High Field (CHiP) at National Central University, the Institute of Physics, Academia Sinica, the National Institute of Physics, University of the Philippines Diliman, as well as Taiwan Semiconductor Research Institute (TSRI).
Y.I. is supported by the National Science and Technology Council (NSTC) of Taiwan under Grant No. 114-2112-M-008-006, and by Academia Sinica under Grant No. AS-TP-112-M01.

\end{acknowledgments}

\bibliographystyle{apsrev4-2}
\bibliography{bibfile}

\end{document}